\documentstyle[aps,12pt,psfig]{revtex}

\newcommand{\be}{\begin{equation}}
\newcommand{\ee}{\end{equation}}
\newcommand{\nn}{\mbox{} \nonumber \\ \mbox{} &&}
\newcommand{\ba}{\begin{eqnarray}}
\newcommand{\ea}{\end{eqnarray}}

\begin{document}
\title{Turbulent 4-wave Interaction of Two Type of Waves}
\author{ Maxim Lyutikov}
\address{Canadian Institute for Theoretical Astrophysics,
 St George Street,
Toronto, Ontario, M5S 3H8, Canada}
\date{\today}
\maketitle

\begin{abstract}
We consider turbulent 4-wave interaction of  two types of waves: 
 acoustic  waves (dispersion $\omega = k$) and 
electromagnetic-type   waves (dispersion $\Omega^2 = m^2 + p^2$). For large wave vectors
($ k  \gg m$), when the dispersion of EM-type waves becomes quasiacoustic,
 the stationary spectra are  obtained  following a standard Zakharov approach.
In the small wave number region ($ k, p \ll m$) we derive nonlinear differential
 Kompaneets-type
kinetic 
equation for arbitrary distributions of the interacting fields
and find when this equation has Kolmogorov-type power law solutions.
\end{abstract}


\section{Introduction}

In this work we consider 4-wave 
interaction of two types of waves: one  with   acoustic dispersion $\omega = k$,
called massless field,
and  another with electromagnetic-type   dispersion $\Omega^2 = m^2 + p^2 $, called massive field
 ($\omega, \Omega$ are frequencies, $k,p$ - wave vectors of the waves, the phase speed 
of the acoustic waves was set to unity).
  This is a fairly common case of wave interaction.
It includes, for example, interaction of electrons with photons and  electromagnetic fields with 
Langmuir waves.
 We assume that the  interacting fields are scalar  fields
obeying  quantum Bose-Einstein  statistics. 
Then the 
collisional integral becomes a nonlinear functional in 
occupation numbers of both  types of waves.  
We are interested in finding stationary solutions of such collision
integral. In particular we will be looking for power law type solutions
($n \propto k^{\alpha}, \, N \propto p^{\beta}$) which can exist in the
inertial ranges $ k , \, p \gg m$ or  $ k , \, p \ll m$. 

A standard procedure to find power law solutions of the kinetic equation
is through Zakharov conformal transformations \cite{Zakharov}. This method
provides a simple rule for finding
stationary power law solutions in the inertial region for differential
scattering (when the change in energy or momentum in each scattering is small).
 For coupled kinetic equations the Zakharov transformation in the general
 cannot be done.
 Tractable interactions include (i) 3-wave interaction
of waves with similar dispersion relations (using  conformal transformations),
  (ii)
3-wave interaction of high frequency and low frequencies waves \cite{Zakharov} 
(using a diffusion approximation). 

In this work we consider  coupled kinetic equations for 4-wave interaction
of waves. 
In the "ultrarelativistic" 
regime, when the wave vectors of both particles is much larger than the mass 
$k\,,p \gg m$,  both types of waves have similar acoustic-type dispersion
and thus can be viewed as selfinteraction between quasiacoustic waves.
The stationary spectra then can be found using conformal transformations.
In the oposite limit of small wave vectors, $k, \,p \ll m$,
different dispersion laws of interacting waves break the homogeneity of the
kernel of the kinetic equation and do not allow exact Kolmogorov-type
solutions. In this case we expand the
collision integral in small changes of energy in each collision.
Then the collision integral may be simply written
as a nonlinear  diffusion equation.
This  method was first applied by
Kompaneets \cite{Kompaneets} who derived a nonlinear equation for the interaction
of the low frequency  boson particles (photons) with nonrelativistic
 classical electrons which were assumed to be in thermodynamic
equilibrium due to quick relaxation by the long range Coulomb forces.

The "ultrarelativistic" regime, when both interacting waves
have quasiacoustic dispersion requires a careful consideration - 
for exactly acoustic dispersion the weak turbulence theory may not be
applicable (e.g., \cite{Balk}- \cite{ZakharovSagdeev}).
Balk \cite{Balk} and Newell and Aucoin \cite{NewellAucoin}  argued that in the multidimensional case
the angular dispersion can justify the applicability of the weak turbulence theory
to acoustic waves, while Zakharov and Sagdeev   \cite{ZakharovSagdeev} resorted to 
small corrections to the acoustic dispersion. We follow the latter approach: in the
limit $p \gg m$ the
correction to the dispersion of EM-type waves will provide the
dispersion required for the final time of interaction of 
acoustic waves and will ensure that kinetic approach is applicable.

The primary purpose of this work is to derive a low frequency  diffusion 
approximation
for  the  4-wave interaction  for arbitrary distribution functions
of both interacting waves  and to find Kolmogorov-type
solutions in the inertial ranges. In case of quantum fields,
 the Kolmogorov-type power law solution may be possible  only in the inertial
ranges,  where energies  of the interacting particles are
 not close to the only quantity with dimension of energy - $m$
and  the occupation numbers satisfy  $n\,, N \gg 1$ or  $n\,, N \ll 1$.

We start with a
Hamiltonian of the system of two interacting waves in terms of wave amplitudes $  a_k$ and
$A_p$   
\ba
&&
{\cal H} = 
\int \omega_k a_k a_k ^{\ast} d^3 k + \int \Omega_p  A_p A_p ^{\ast} d^3p +
\nn
{1\over 4} 
\int T_{kpk'p'} a_k ^{\ast} A_p ^{\ast}  a_{k'}  A_{p'} 
\delta({\bf k} + {\bf p} - {\bf k}' - {\bf p}') d^3 k 
d^3 p  d^3 k'  
 d^3 p' 
\label{H}
\ea

We assume the simplest form of the  relativistically invariant matrix element 
(e.g., \cite{LandauIV})
\be 
T_{kpk'p'} = { g^2  \over \omega_k \Omega_p \omega_{k'} \Omega_{p'}}
\ee
Then the kinetic equations for  occupation numbers $n_k, N_p$ are
\be
\left( 
\begin{array} {c}
{\partial  n_k \over \partial t} \\
{\partial  N_p \over \partial t}
\end {array} \right)=
{\pi \over 2} \int T^2 _{kpk'p'}  
 F[n,N]
\delta(\omega_k + \Omega_p-\omega_{k'} - \Omega_{p'})
\delta({\bf k} + {\bf p} - {\bf k}' - {\bf p}')
\left( 
\begin{array} {c}  d^3 p \\
d^3 k \end {array} \right) d^3 k' d^3 p'
\label{kk}
\ee
where
\be
F[n,N]=  N_{p'} n_{k'} \left( n_k +   N_p +1 \right) -
 n_k  N_p  \left(   N_{p'} +  n_{k'} \right)
\label{dq}
\ee

\section{Applicability of the kinetic equation}

For the applicability of the weak turbulence theory (based on random phase approximation)
we need that
the typical interaction time 
$
 t_{int}  \approx  1/(T  N_{tot})$, (where $N_{tot}=\int d^3 p N_p$ and
$T$ is a matrix element for amplitudes)
be much larger than chaotization time for the phases of waves $ t_{collision}= 1/ \Delta k v_g$
($v_g $ is the group velocity and the wave packet is centered on $k_0 \pm \Delta k$)
and diffusion time  $ t_{diff} =  1/ \Delta k^2 v_g^{\prime}$.

For a power law $\omega \sim  k^{\alpha}$ we have $v_g^{\prime} \sim
\alpha (\alpha-1)k^{\alpha -2}$, so
$t_{collision} \sim  { k \over  \Delta k} {1 \over \alpha \omega }$ which is 
can be very short - of the order $1/\omega$. On the other hand the diffusion time
 $ t_{diff} = ( \Delta k/k)^{-2} / ( \alpha (\alpha-1) \omega).$ 
Which obviously unapplicable to acoustic turbulence. Still, 
if $\alpha \neq 1$, then the diffusion time 
$ t_{diff}$ also can be very short. This case (nonacoustic turbulence)
is applicable for small wave vectors: $k \ll m$.
\begin{equation}
{ t_{diff} \over t_int}  =  { T  N  / k c } \sim  { g^2 n k \over m} 
\end{equation}
 where we estimated $T (k \ll m) = g^2  /( k m )$ , $ N= n k^3 $.
Thus the applicability of the kinetic equation in this case requires
\begin{equation}
 n  \ll   {  m \over  g^2 k} 
\end{equation}
which for $k \leq m$ reducesto $n \ll   {1 \over  g^2}$.

On the other hand for 
semiacoustic waves with $k \gg m$ we find
 $  \omega = k\, c ( 1 +  { m^2  \over   k ^2}  )$ we have
$\omega^{\prime \prime}= { m^2 \over    k^3 } $, 
$ t_{diff} = \left( {k \over \Delta k} \right)^{2} { k  \over  m^2 }
\approx{ k  \over  m^2}
 $,
then the applicability criterion is
\begin{equation}
{ t_{diff} \over t_{int}} 
 = { T  N    k \over   m^2  } =  {  g^2 n   k^2 \over   m^2   }
 \ll 1
\end{equation}
(the matrix element in this limit is
 $ T = g^2/k^2 $).
This will eventually break at 
\begin{equation}
k_{max} =  { m \over g \sqrt{n} }  
\end{equation}
Thus for applicability of a kinetic equation in a range $m\ll k \ll k_{max}$ 
 the same condition should be satisfied, $ n \ll {1 \over g^2}  $.

We can also estimate when the higher 
 terms will be unimportant in the kinetic equation.
The ratio of the second  order terms of expansion  in $g^2$ to the first order 
is typically
\begin{equation}
{[2]/[1]} =  { g^4 k^3 n^2 / \omega_1  \omega_2 \omega  }
\end{equation}
For acoustic turbulence  $k \gg m$ the frequencies are $\omega_i\sim k$, this gives
\begin{equation}
{[2]/[1]}_a =  { g^4  n^2} \mbox{ $ \ll 1 $ if $ 
n \ll {1 \over  g ^2 } $}
\end{equation}
while for small wave vectors 
$\omega_1=m$ $\omega_2=k$
\begin{equation}
{[2]/[1]}=  { g^4  n^2 k / m}  \mbox{ $ \ll 1 $ if $ n \ll {1 \over  g ^2 } $}
\end{equation}

Thus we conclude that 
for existence of inertial range where weak turbulence theory is
applicable the occupation numbers should satisfy
\be
 N _p, n_k\ll {1 \over g^2}  
\label{t}
\ee

\section{ Interaction of particles at large wave vectors }

In the limit $k,p \gg m$ ("relativistic particles") the dispersion law
for the EM-type  waves become almost acoustic
\be 
\Omega \approx k \left( 1 + { m^2 \over 2 k^2} \right)
\ee
The applicability of the weak turbulence theory  to the  interaction of acoustic
wave  in more than one dimension is a long standing question. 
In 1-D weak turbulence theory is not applicable since
the interaction time between two waves is infinite, 
so that mutual influence of waves becomes large.
In larger dimensions there were claims  (\cite{NewellAucoin}, \cite{Balk}) 
that angular dispersion 
 may limit the interaction time sufficiently
to allow weak turbulence approximation. Another approach is to
allow for small corrections to the dispersion which will 
  limit the interaction time
of wave packets and allows us to implement methods 
of the weak turbulence theory regardless of angular dispersion \cite{ZakharovSagdeev}.

Assuming the weak turbulence theory is applicable (Eq. (\ref{t})) we can 
find  stationary solutions to the kinetic equation using Zakharov conformal transformations.
In the limit $k,p \gg m$ the dispersion laws $ \omega \propto k ^{\alpha}$ and $\Omega \propto p^{\alpha}$
with $\alpha=1 $, and  the interaction coefficient in this limit is a homogeneous function
of the order $-2$:
\be 
T_{kpk'p'}( \lambda {\bf k}, \lambda {\bf p}, \lambda {\bf k'}, \lambda {\bf p'}) =
\lambda^p T _{k123}( {\bf k},  {\bf p},  {\bf k'},  {\bf p'}),
\hskip .2 truein p =-2
\ee
Zakharov transformations then gives stationary solutions

\be
n_P ,\, N_P \propto \omega ^{ -5/3},
\hskip .2 truein
n_Q  ,\,  N_Q \propto \omega ^{ -4/3}
\label{PQ}
\ee 
Here $P $ and $Q$ symbolize Kolmogorov spectra with constant flux of energy and
constant flux of action. 
It is possible to check that collision integral converges for both spectra, so 
that the interaction is local.

The sign  of fluxes $\mbox{sign}[ P] > 0$ and 
$\mbox{sign} [Q]  <0$ imply that two possible stationary spectra include the
constant   energy flux  flowing to large $k$ or
constant  flux of   particles to small $k$.
Realization of a particular spectra (Eqns (\ref{PQ})
depends on the boundary condition at the limits of the inertial range, e.i., by the
location of the sinks and sources of energy and action. 
If the source  of energy and  of  action are not at $k=0 $  and $k=\infty$ correspondingly, then,
depending on values of $p$ and $\alpha$, the integrals of energy or action
converge or diverge. 
Using criteria for the stability of Kolmogorov spectra \cite{Zhakharovbasic}, 
 we conclude that for sources located at $k_0,p_0 \gg m$ it is 
{\it  Kolmogorov-in-action}  case that is realized
 (the action integral  diverges  at large $k$).  In this case the turbulence
is Kolmogorov in action - particles are  flowing to small k (and may eventually be 
 stored in Bose condensate). For  $k,p < k_0, p_0$  a stationary   Kolmogorov-in-action
 spectrum will form, while for
for $k,p > k_0, p_0$  there 
will be  a selfsimilar  "thermal wave"  type solution carrying energy  to large
$k$ and $p$:
\begin{equation}
n_k = { 1\over t^{ {3 \over 7}} } n_0 ( { \omega \over  t^{ {5\over 7}} }),
\hskip .2 truein \mbox{for $ k > k_0$}
\end{equation}
where it is eventually dissipated \cite{Zhakharovbasic}.
Thus, in the "thermal wave" 
  the  mean frequency increases with time  $\propto t^{ {5\over 7}}$.

This situation (flow of particles to small wave numbers, where stationary spectrum is
formed,  and flow of
energy to large  wave numbers)  is somewhat similar to interaction of Langmuir waves where
in the low frequency region the dominant nonlinear process
for Langmiur waves is the exchange of virtual ion sound waves which
leads accumulation of particles with small $k$ - Langmiur collapse.

\section{ Interaction of particles at  small wave vectors}

\subsection{Kompaneets approximation}

In the limit $k,p  \ll  m$ the change in energy in each collision is small, so we
can expand the collision integral in small changes of the energies of particles:
\begin{eqnarray}
&&
\omega_{k'} = \omega_k +\Delta
\nonumber \\
&&
\Omega_{p'}=\Omega_p- \Delta
\end{eqnarray}
where $\Delta $ to the first order in 
$\omega/m$ is given by
\begin{equation}
\Delta =
-{{\omega\,\left( \omega\,\left( 1 - {\bf n}\cdot {\bf n'} \right)  + 
        {\bf p}\cdot \left( {\bf n} - {\bf n'} \right)  \right) }\over 
    {m + \omega\,\left( 1 - {\bf n}\cdot {\bf n'} \right)  - 
      {\bf p}\cdot {\bf n'}}}
\end{equation}
Expansion of $F[n,N]$ to the second order in $\Delta$  gives
\begin{eqnarray}
&&
F[n,N] =\Delta \left(   N( N+1) { \partial n \over \partial  \omega} -
n(n+1) { \partial  N \over \partial  \omega} \right) +
\nonumber \\
&&
{ \Delta^2  \over 2} \left(   N( N+1) { \partial ^2 n \over \partial  \omega^2} +
n(n+1) { \partial^2   N \over \partial  \omega^2} - 2
{ \partial  N \over \partial  \omega}  { \partial n \over \partial  \omega} 
\left( n+ N+1 \right) \right)
\label{s2}
\end{eqnarray}

In the following we concentrate on the kinetic equation for particles $n$.
In the kinetic equations we next separate angular  and momentum integrations
\ba
&&
I_{coll}(k) = 4\pi \sigma_0
\int _0^{\infty} p^2 dp 
\left[  \left\{  
\int  { d{\bf \Omega_{p}}  \over  4\pi}{   d{\bf \Omega_{n'} } \over  4\pi}
 \Delta( ( k, {\bf p},{\bf n}' )^2 \right\}
 \left( {\partial N \over \partial  \Omega } n ( n +1) 
- {\partial n  \over \partial  \omega }  N (  N+1) \right)
\right.
\nn
\left.
+ {1\over 2}\left\{
 \int {  d{\bf \Omega_{p}} \over  4\pi} {   d{\bf \Omega_{n'} } \over  4\pi}
 \Delta( ( k, {\bf p},{\bf n}' )^2 \right\}
\left( n ( n +1) {\partial^2  N \over \partial  \Omega^2}+
 N (  N+1)  {\partial ^2 n  \over \partial ^2 \omega } -
2 {\partial N \over \partial  \Omega }  {\partial n  \over \partial  \omega } 
\left( N+ n +1 \right) \right) \right]
\ea
where
\be
\sigma_0 =  { 1\over 16 \pi} { \left| M_{fi}\right|^2 \over m^2} 
\ee
Denoting the angle averages
\begin{eqnarray}
&&
I_1 (p,k)= \sigma_0
 \int { d{\bf \Omega_{p}}   \over 4 \pi} { d{\bf \Omega_{n'} }   \over 4 \pi} 
\Delta( ( k,{\bf p}, {\bf n}') =
{{k\,\left( 4\,{k^2} - 3\,k\,m + {p^2} \right) }\over 
   {3\,{m^2}}} \sigma_0
\nn 
I_2(p,k)=  \sigma_0
\int {  d{\bf \Omega_{p}}   \over 4 \pi} {    d{\bf \Omega_{n'} }   \over 4 \pi} 
 \Delta( ( k, {\bf p},{\bf n}' )^2=
{{2\,{k^2}\,\left( 2\,{k^2} + {p^2} \right) }\over 
   {3\,{m^2}}} \sigma_0
\ea
We can rewrite collision integral in the form
\ba
&& I_{coll}(k) = 
4\pi \sigma_0
 \int _0^{\infty} p^2 dp
\left[ 
I_1 (p,k) \left( {\partial N \over \partial  \Omega } n ( n +1)
- {\partial n  \over \partial  \omega }  N (  N+1) \right) \right.
\nn
\left.
+ {1\over 2} I_2 (p,k)\left( n ( n +1) {\partial^2  N \over \partial  \Omega^2}+
 N (  N+1)  {\partial ^2 n  \over \partial ^2 \omega } -
2 {\partial N \over \partial  \Omega }  {\partial n  \over \partial  \omega }
\left( N+ n +1 \right) \right) \right]
\label{Pw}
\ea

There  is a also a constraint condition on the equation (\ref{Pw}) - it sould conserve total
number of particles, which is an integral of motion of the Hamiltonian (Eq. \ref{H}):
\be
\int d^3 k I_{\rm coll} (k) =0
\ee 
 Partially integrating and 
collecting terms with different powers of distribution function we find 
\ba
&&
\int d \omega d \Omega  N n \left[
\left( {  \partial   \over \partial  \Omega} -  {  \partial   \over \partial   \omega} \right)
i_2 +
{1\over 2} \left( {  \partial^2 \over \partial^2 \Omega} +
 {  \partial^2 \over \partial^2 \omega} -
2 {  \partial^2 \over \partial \Omega \partial \omega } \right) i_2 \right] +
\nn
\int d \omega d \Omega  N^2 n {  \partial   \over \partial   \omega} 
\left[
 - i_1 + {1\over 2} \left( {  \partial   \over \partial   \omega} -
 {  \partial   \over \partial  \Omega} \right)  i_2 \right] +
\nn
\int d \omega d \Omega  n ^2 N  {  \partial   \over \partial   \Omega}
\left[
i_1 - {1\over 2} \left( {  \partial   \over \partial   \omega} -
 {  \partial   \over \partial  \Omega} \right)  i_2 \right] =0
\label{s8}
\ea
where
\be
i_1 = k^2 \, {\partial k \over \partial \omega} \, p^2  \, {\partial p \over \partial \Omega}  \, I_1 
, \hskip .2 truein 
i_2=  k^2  \, {\partial k \over \partial \omega}  \, p^2  \, {\partial p \over \partial \Omega}  \,I_2  
\ee
Each term in Eq. (\ref{s8}) should be equal to zero. This gives a relation between the 
coefficients $i_1$ and $i_2$:
\be
i_1 = {1\over 2} \left( {  \partial i_2  \over \partial \Omega} -
{  \partial i_2 \over \partial \omega} \right)
\label{s}
\ee
This is an important relation that is one of the main results of the work. It relates the
coefficients in the diffusion expansion of the collisional integral independently of the
particular form of the interaction. 

Using  relation  (\ref{s}) we can 
simplify  considerably the collision integral (Eq. \ref{kk}). Partially integrating we find that
collision integrals take a particularly simple diffusion type form:
\ba
&&
I_{\rm coll} (k) = 
 2\pi \sigma_0 \int p^2 dp  {  \partial \over   \partial    \omega}
\left( \omega^2 I_2 \left[  n'  N (N+1) -  N' n (n+1) \right] \right) 
\nn 
I_{\rm coll} (p)= 
 2\pi \sigma_0
\int  k^2  d k {  \partial \over   \partial     \Omega }
\left( \sqrt{ \Omega } I_2  \left[ n'  N (N+1) -  N' n (n+1) \right] \right)
\label{int2}
\ea
For massless particles $n$ and massive particles $N$ correspondingly.

Equations (\ref{int2})  are the diffusion-type equations for the low frequency interaction
of two types of quantum  waves  - "massless" acoustic waves "n" and
"massive" EM waves $N$. 
To reduce then to the more familiar Focker-Plank form we  introduce coefficients
\ba
&
A_1= 4 \pi \int p^2 dp N(N+1),
 & A_2 =  4 \pi \int k^2 dk { k^2  } n (n+1)
\nonumber \\ &
\omega_o^2 = {1 \over 2} {  \int p^2 d p N (N+1) p^2  \over 
 \int p^2 d p N (N+1) } ,  &
\Omega_0 =
{ \int k^2 dk { k^4  \over 4 m} n (n+1)
\over \int k^2 dk { k^2  } n (n+1) }
\nonumber \\ &
B_1 =- {  \int p^2 d p N '  p^2  \over  \int p^2 d p N (N+1) p^2 } ,
& B_2 = - {  \int k^2 dk { k^2  } n'
  \over \int k^2 dk { k^2  } n (n+1) }
\nonumber \\ &
C_1 =  {  \int p^2 d p N '  p^2  \over \int p^2 d p N ' } 
{ \int p^2 d p N (N+1)  \over \int p^2 d p N (N+1) p^2 } -1  ,
& C_2 = { \int k^2 dk { k^4  \over 4 m} n'
\over \int k^2 dk { k^2  } n' } 
{\int k^2 dk { k^2  } n (n+1) \over
\int k^2 dk { k^4  \over 4 m} n (n+1) }-1
\label{S0}
\ea 

Note,  that in the thermodynamic equilibrium 
$N' =-N(N+1)/T $, so that $B_1, B_2 = 1/T$ (T is  equilibrium temperature) and $C_1, C_2=0$.
Coefficients $A_{1,2}$ and $\omega $, $\Omega$ are complicated even for thermodynamic 
equilibrium. In the classical limit they give $A_{1,2}=  N_{tot}, \, n_{tot}$ - total number of particles,
$\omega_o^2 = 3 m T$ -average $<k^2>$ and $\Omega_0 = 5T/4$.

Using notations (\ref{S0}) the collision integrals may be written as
\ba 
&&
I_{\rm coll} (\omega) = { \sigma_0 \over 2}  A _1
\left[ \omega^4 ( \omega^2 +  \omega_0^2) \left(  {\partial  n  \over \partial \omega} +
B_1 n (n+1) \left(1 +{  \omega_0^2 C_1 \over 
\omega^2 +  \omega_0^2 }  \right)  \right) \right]
\nn
I_{\rm coll} (\Omega) = { \sigma_0 \over 2} A _2
\left[\sqrt{ \Omega}  ( \Omega +  \Omega_0) \left(  {\partial  N  \over \partial \Omega} +
B_2 N (N+1) \left( 1+ {  \Omega_0 C_2 \over 
\Omega +  \Omega_0 } \right) \right) \right] 
\label{w}
\ea

\subsection{Stationary solutions}

To find  stationary power law solutions 
 to the nonlinear equations (\ref{w}) we  assume that $C_{1,2} \approx 0$ (in this case the
resulting solution  are sums of power laws (see Eqns. (\ref{wq}-\ref{wq1})). In case
of $C_1$ the dominant contribution  to the number density comes from the short wave vector limit of the
thermodynamic equilibrium (see Eq. (\ref{wq1})), for which we can assume  $C_1 \approx 0$. 
The coefficient $C_2 $ on the other
hand is generally nonzero, but if the  maximal energy of particle $n$ (cutoff of the power law)
is much less than the mass $m$ then the term with $C_2 $ is approximately 
$p_{max}/m$  - much smaller than unity.
 
Under assumption $C_{1,2} \approx 0$
the  Kompaneets-type equations for the occupation numbers $n$ and $N$ become
\ba
&&
 {\partial n \over \partial t} = 
-  { \sigma_0 \over 2} A_1{1\over \omega^2}  {\partial  \over \partial \omega}
\left[ \omega^4 ( \omega^2 +  \omega_0^2) \left(  {\partial  n  \over \partial \omega} +
B_1 n (n+1) \right) \right]
\nn
 {\partial N \over \partial t} = - { \sigma_0 \over 2} A_2
 {1\over  \sqrt{\Omega} }   {\partial  \over \partial \Omega}
\left[ \sqrt{\Omega}  ( \Omega + \Omega _0) \left(  {\partial N \over \partial \Omega} +
B_2 N (N+1)  \right) \right]
\label{S}
\ea
and the 
 the  stationary equations ${\partial n \over \partial t} = {\partial N \over \partial t} = 0$
should satisfy 
\ba
&&
\omega^4 ( \omega^2 +  \omega_0^2) \left( n' +B_1  n(n+1) \right) = P_1
\nn
\sqrt{ \Omega} (  \Omega + \Omega_0) \left(N'  +B_2 N(N+1) \right) = P_2
\label{S1}
\ea
The physical meaning of the constants $P_1$ and $P_2$ is the total particle flux in the wave vector
space.

Eqns (\ref{S1}) are nonlinear Riccatti-type differential equations. Using a conventional substitution
$N' (x) \rightarrow U' /U $ they can be reduce to linear equations of the form
\be
  U''  + U' - G(x) U=0
\ee
It is not generally integrable in quadratures, but in some limits it may have a solution
which is a sum of power laws \cite{Kamke}.
\footnote{For the  equilibrium distribution of heavy particle $N$  various approximations
for $n$ in the limit
$\omega \gg \omega _0$ and $N\ll 1$ are summarized  by  \cite{SunyaevTitarchuk}.} 
The best way to search for possible power law solutions is to use the substitution
$ U (x) \rightarrow W \exp \{ \left( \int \sqrt{G(x)}  \right) ^{-1} dx \} $
 which  takes a correct account of the leading  power law term. 

Power law solutions of the equation (\ref{S1}) are possible in the inertial regions - where there is no 
quantity with a dimension of length: these are regions $ \omega $ much less or much larger
than $  \omega _0$, 
$ \Omega  $ much less or much larger
than $ \Omega _0$ and $ N,n $ much less or much larger
than $1$.

In the linear regime $N,n \ll 1$ the stationary solutions are 
\ba
&&
n \propto \int { P _1\over \omega^3 (\omega^2 + \omega_0^2)} d \omega
\nn
N \propto \int { P_2  \sqrt{\Omega}  \over \
 (\Omega + \omega_0)} d \Omega
\ea

In the strongly nonlinear ($ N,n \gg 1$)
 limit
the leading  orders in power-laws are
\ba
&&
n_{
\omega}  \propto  \left\{
\begin{array}{ll}
{ P _1 \over \omega_0 \omega^2} + {1\over B \, \omega} & \mbox{ for $\omega \ll \omega_0$} \\
 {P _1 \over \omega^3} + {3\over 2 B\, \omega} & \mbox{ for $\omega \gg \omega_0$} 
\end {array} \right.
\label{wq}
\\ &&
N_{\Omega} \propto \left\{
\begin{array}{ll}
 {1\over  B \Omega}  - {P_2\over  \Omega_0^{1/2} \Omega^{1/4}} & \mbox{ for $ \Omega \ll \Omega_0$} \\
 {1\over  B \Omega}  - {P_2\over \Omega^{3/4}} & \mbox{ for $ \Omega \gg \Omega_0$}
\end {array} \right.
\label{wq1}
\ea
Given the solutions (\ref{wq1}), 
we can also calculate the coefficients $\Omega_0$ and $\omega_0$:
\ba
&&
\Omega_0 =
\left\{ \begin{array}{cc}
{k_{max}^2 \over 8 m}& \mbox{for $n \propto \omega^{-2}$} \\
{k_{max}^2 \over 12 m}& \mbox{for $n \propto \omega^{-3}$} 
\end{array} \right.
\nn
\omega_0
= p_{max}/\sqrt{3}
\label{wq2}
\ea
where $k_{max}$ and $p_{max}$ are  the upper cut-offs limits for the corresponding power laws.

We can verify that the integral in the expressions
for the coefficients (Eq. (\ref{S0}) converge at $k,p=0$. Note also, that
for massive particles $N$,  the leading term in the  small wave vector limit 
corresponds to the low frequency asymptotics of the thermodynamic equilibrium 
spectrum.  

Equations  (\ref{wq}-\ref{wq2}) give the selfconsistent Kolmogorov-type solutions 
to turbulent interaction of two types of waves. They are qualitatively shown in Fig. (\ref{Fig}).
In all cases the particles are flowing to small $k,p$ with a constant flux in phase space.
If the sources of particles are located at $k,p \gg m$ (Fig.  \ref{Fig} (a)),
 so that $\omega_0 \approx m$, then
the low energy tail  of the distribution of massless particles ($\omega \ll \omega_0$)
  is $n \propto \omega^{-2}$. Alternatively, if the sources of particles are located at $k,p \ll m$
(Fig.  \ref{Fig} (b)), then
the low energy tail  of the distribution of massless particles 
 is $n \propto \omega^{-3}$.

\section{Conclusion}

We have considered turbulent 4-wave interaction of two types of 
 waves with different dispersion laws:
acoustic and electromagnetic-type. Both types of waves were assumed to obey 
a quantum boson
statistics. We 
 found possible stationary Kolmogorov-type spectra in the inertial ranges 
$k, \, p \gg m$ and $k, \, p \ll m$. To reach the stationary solutions the
  particles  should be generated at finite momenta $k$ and $p$ and
flow to small wave numbers. 
In the "ultrarelativistic limit"  $k, \, p \gg m$, the  turbulent interaction of two
types of waves resembles self interaction of semiacoustic waves with the
dispersive correction ensuring the applicability of the weak turbulence limit. 
In the small wave vector limit $k, \, p \ll m$ the collisional integral may be
reduced to diffusion-type differential equation. 
In this limit, the
 small wave number asymptotic of massive particles is dominated
by the thermal equilibrium tail $N \propto \Omega^{-1}$, while the small wave number asymptotic of
massless particles is steeper: $n \propto \omega ^{-2} \div \omega ^{-3}$. Arguably, the 
massless particles $n$  will eventually reach $k=0$ to be 
absorbed in the Bose condensate.\footnote{Limitations of the
 kinetic approach to Bose condensation were discussed by Kogan et al. (1992)}

I would like to thank Alexander Balk for useful comments.

\newpage

FIGURE CAPTIONS

FIG. I. 
Stationary power law solutions of the system (\ref{t}). (a) source of particles is located
at $k,p \gg m$, (b) source of particles is located at $k,p \ll  m$

\newpage

\begin{figure}
\psfig{file=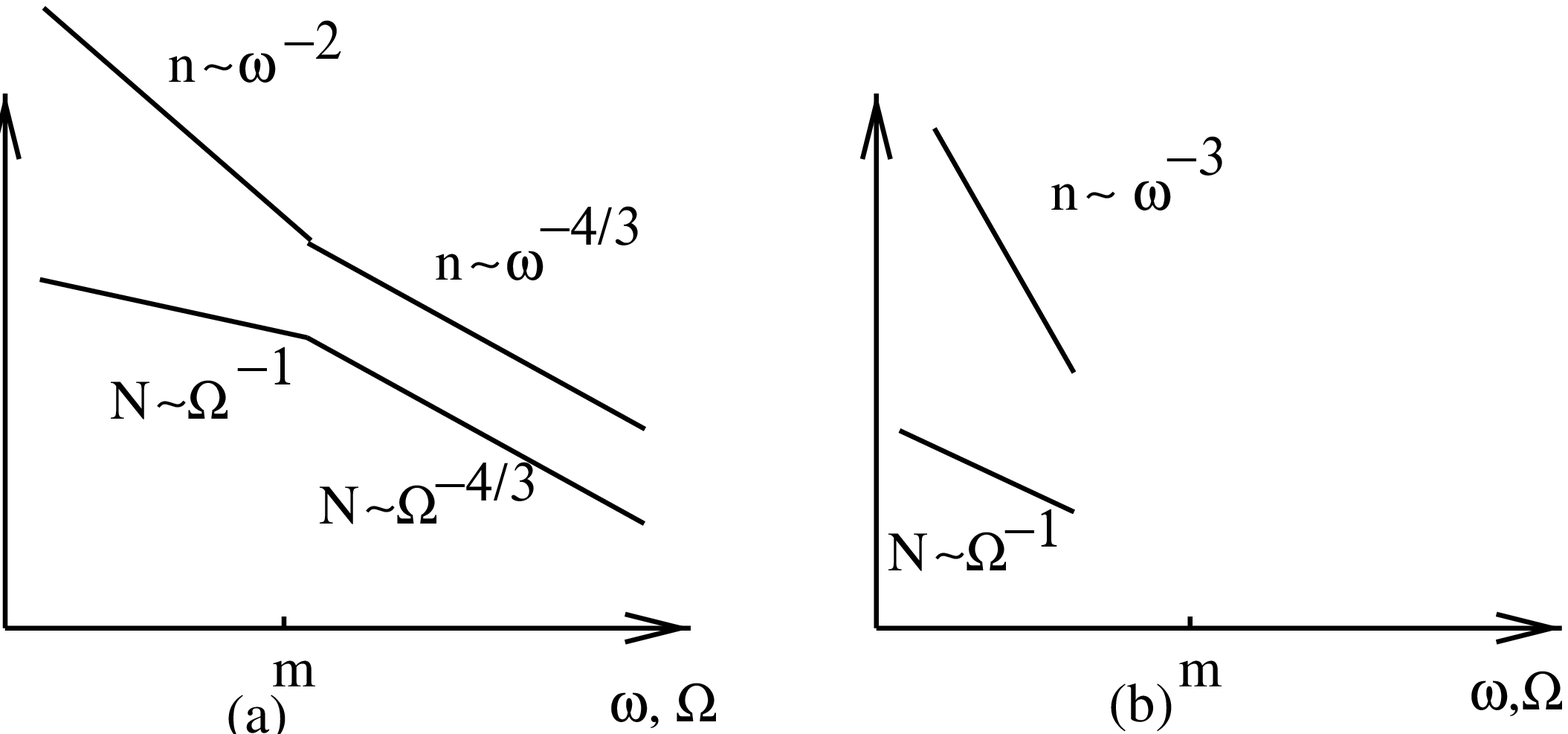,width=15cm}
\label{Fig}
\end{figure}
\end{document}